\documentstyle[referee,psfig,epsf]{l-aa}
\def\lsim{\lower.5ex\hbox{$\; \buildrel < \over \sim \;$}}
\def\gsim{\lower.5ex\hbox{$\; \buildrel > \over \sim \;$}}
\newcommand{\eqb}{\begin{eqnarray}}
\newcommand{\eqe}{\end{eqnarray}}

\newbox\grsign \setbox\grsign=\hbox{$>$} \newdimen\grdimen \grdimen=\ht\grsign
\newbox\simlessbox \newbox\simgreatbox
\setbox\simgreatbox=\hbox{\raise.5ex\hbox{$>$}\llap
   {\lower.5ex\hbox{$\sim$}}}\ht1=\grdimen\dp1=0pt
\setbox\simlessbox=\hbox{\raise.5ex\hbox{$<$}\llap
   {\lower.5ex\hbox{$\sim$}}}\ht2=\grdimen\dp2=0pt
\def\gsim{\mathrel{\copy\simgreatbox}}
\def\lsim{\mathrel{\copy\simlessbox}}

\def\ref{\par\noindent\hangindent=1.5cm}
\begin{document}
\thesaurus{02.01.2, 02.02.1, 02.19.1, 08.14.1, 08.09.2:SS433, Section 06}
 
\title{Estimation and Effects of the mass outflow rate from shock compressed flow around compact objects }
 
\author{Sandip K. Chakrabarti}
\institute{S.N. Bose National Centre for Basic Sciences,
JD-Block, Sector III, Salt Lake, Calcutta 700091, India}

\date{Received \dots ; accepted \dots}
\maketitle
\markboth{S.K. Chakrabarti: Outflow from Shocked Disks}{}

\maketitle

\begin{abstract}

Outflows are common in many astrophysical systems which 
contain black holes and neutron stars. Difference between 
stellar outflows and outflows from these systems is that 
the outflows in these systems have to form out of the 
inflowing material only. The inflowing material can 
form a hot and dense cloud surrounding the compact object, 
either because of a centrifugal barrier, or a denser barrier 
due to pair plasma or pre-heating effects. This barrier 
behaves like a stellar surface as far as the mass loss is 
concerned. We estimate the outflow rate from the regions of 
shock compressed flow. The outflow rate is directly
related to the compression ratio of the gas at the shocks.
These estimated rates roughly match the rates in real observations 
as well as those obtained from numerical experiments. In special geometries,
where the solid angle of the outflow is higher, the disk evacuation
takes place creating quiescence states. Outflows are shown to be important in deciding the
spectral states and Quasi Periodic Oscillations (QPO)s of observed X-rays.

\end{abstract}


\noindent Accepted for Publication in Astronomy and Astrophysics (July 1999)

\section{Introduction}
 
Cosmic radio jets are believed to originate from the
centers of active galaxies which harbor black holes 
(e.g., Chakrabarti 1996, hereafter C96). Even in so-called 
`micro-quasars', such as GRS 1915+105 which are believed to have
stellar mass black holes (Mirabel \& Rodriguez 1994), the outflows are common. 
The well collimated outflow in SS433 has been well known for almost two decades (Margon 1984).
Similarly, systems with neutron stars also show outflows, as is 
believed to be the case in X-ray bursters (e.g. Titarchuk 1994).

There is a large number of articles in the literature
which attempt to explain the origin of these outflows. These articles 
can be broadly divided into three sets. In one set, the jets
are believed to come out due to hydrodynamic or magneto-hydrodynamic 
pressure effects and are treated separately from the disks 
(e.g., Fukue 1982; Chakrabarti 1986). In another set, efforts are made to 
correlate the disk structure with that of the
outflow (e.g., K\"onigl 1989; Chakrabarti \& Bhaskaran 1992). 
In the third set, numerical simulations are carried out
to actually see how matter is deflected from the equatorial plane 
towards the axis (e.g., Hawley 1984; Eggum et al. 1985; Molteni et al.
1994). Nevertheless,
the definitive understanding of the formation of outflows is still
lacking, and more importantly, it has always been difficult to
estimate the outflow rate from first principles. In the first set, 
the outflow is not self-consistently derived from the inflow. 
In the second set, only self-similar steady solutions are 
found and in the third set, either a Keplerian
disk or a constant angular momentum disk was studied, neither being
the best possible assumption. On the other hand, the mass outflow
rates of the normal stars are calculated very accurately from the 
stellar luminosity. The theory of radiatively driven winds seems to be
very well understood (e.g. Castor et al. 1975). Given that
the black holes and the neutron stars are much simpler celestial objects, 
and the flow around them is sufficiently hot to be 
generally ionized, it should have been simpler to compute the 
outflow rate from an inflow rate by using the methods employed 
in stellar physics.

Our approach to the mass outflow rate computation is somewhat different
from that used in the literature so far. Though we consider simple
minded equations to make our points, such as those applicable 
to conical inflows and outflows, we add a fundamental ingredient 
to the system, whose importance is being revealed only very recently 
in the literature. This is the quasi-spherical centrifugally 
supported dense atmosphere with a typical size of a few tens of  Schwarzschild 
radius around a black hole and a neutron star. Whether a shock
actually forms or not, this dense region exists, as long as the 
angular momentum of the flow close to the compact object is roughly
constant and is generally away from a Keplerian distribution as 
is the case in reality (C96). This centrifugally 
supported region (which basically forms the boundary layer of 
black holes and weakly magnetized neutron stars) successfully
replaced the so called `Compton cloud' (Chakrabarti \& Titarchuk 1995
[hereafter CT95]; Chakrabarti et al. 1996) 
in explaining hard and soft states of black hole
candidates, and the converging flow property of this region successfully
produced the power-law spectral slope in the soft states of black hole
candidates (CT95).
The oscillation of this region successfully explains the general
properties of the quasi-periodic oscillation (Molteni et al., 1996,  C96) of X-rays from black holes
and neutron stars. It is therefore of interest to know if this region plays 
any major role in the formation of outflows.

Several authors have also mentioned denser regions forming due to different physical effects.
Chang \& Ostriker (1985) showed that pre-heating of the gas could produce standing
shocks at a large distance.  Kazanas \& Ellison (1986) mentioned 
that pressure due to pair plasma could produce standing shocks at smaller distances 
around a black hole as well. Our computation is insensitive to the actual 
mechanism by which the boundary layer is produced. All we require is that the gas should
be hot in the region where the compression takes place (i.e., the optical 
depth should ne small). Thus, since Comptonization
processes cool this region (CT95) for larger accretion rates (${\dot M} \gsim 0.1 
{\dot M_{Eddington}}$) our process will produce outflows in hard states and low luminosity objects
consistent with current observations. Some workers talked about 
a so-called `cauldron' model of compact objects where jets were assumed to
emerge from a dense mixture of matter and radiation through a de-Laval nozzle as in the `twin-exhaust' 
model (for a review of these models see Begelman et al. 1984). The difference 
between this model and the present one is that there a very high accretion 
rate was required (${\dot M}_{in} \sim 1000 {\dot M}_E$)  while we consider thermally driven outflows
from smaller accretion rates. Second, the size of the `cauldron' was thousands of
Schwarzschild radii (where gravity was so weak that the channel has to have the shape of
a de-Laval nozzle), while we have a CENBOL of about $10 R_g$ (where the gravity
plays an active role in creating the transonic wind) in our mind. 
Third, in the present case, matter is assumed to pass through a sonic 
point using the pre-determined funnel where rotating pre-jet matter is 
accelerated (Chakrabarti 1984) and not through a `bored nozzle'
even though symbolically a quasi-spherical CENBOL is considered for mathematical convenience.
Fourth, for the first time we compute the outflow rate completely analytically starting from the
inflow rate alone. To our knowledge such a calculation has not been done in the literature at all.

Once the presence of our centrifugal pressure supported boundary layer (CENBOL) 
is accepted, the mechanism of the formation of the outflow becomes clearer. One basic criterion
is that the outflowing winds should have a positive Bernoulli constant (C96) (although
in the presence of radiative momentum deposition, a flow  with negative initial energy
could also escape as outflow, see Chattopadhyay \& Chakrabarti, 1999). Just as photons
from the stellar surface deposit momentum on the outflowing wind and keep the flow
roughly isothermal at least up to the sonic point, one may assume 
that the outflowing wind close to the black hole is kept isothermal due to 
deposition of momentum from hard photons. 
In the case of the sun, it's luminosity is only $10^{-5}\ L_{Edd}$ and the typical mass outflow
rate from the solar surface is $10^{-14}M_\odot$ year$^{-1}$ (Priest, 1982). Proportionately, for a
star with an Eddington luminosity, the outflow rate would be $10^{-9} M_\odot$ year$^{-1}$. This is
roughly half the Eddington rate for a stellar mass star. Thus, if the flow is 
compressed and heated at the centrifugal barrier around a black hole, it would also
radiate enough to keep the flow isothermal (at least up to the sonic point) if the efficiency
were exactly identical. Physically, both requirements may be equally
difficult to meet, but in reality with photons shining on outflows near a black hole with almost
$4\pi$ solid angle (from the funnel wall) it is easier to maintain the isothermality in the slowly moving
(subsonic) region in the present context. Another reason is  
this: the process of momentum deposition on electrons is more efficient near a black hole.
The electron density $n_e$ falls off as $r^{-3/2}$ while the photon density $n_\gamma$
falls off as $r^{-2}$.Thus the ratio $n_e/n_\gamma \propto r^{1/2}$ increases with the size of the region.
Thus a compact object will have a lesser number of electrons per photon and the momentum transfer is
more efficient. In a simpler minded way, the physics is scale-invariant, though. In solar physics, it is
customary to choose a momentum deposition term which keeps the flow isothermal to be of the
form (Kopp \& Holzer, 1976; Chattopadhyay \& Chakrabarti, 1999),
$$
F_r = \int_{R_s}^r D dr,
$$
where $D$ is the momentum deposition (localized around $r_p$) factor with a typical spatial dependence,
$$
D=D_0 e^{-\alpha (r/r_p-1)^2}.
$$ 
Here, $D_0$, $\alpha$ are constants and $R_s$ is the location of the stellar surface.
Since $r$ and $r_p$ appear in ratio, exactly the same physical consideration would be
applicable to black hole physics, with the same result {\it provided} $D_0$ is scaled with
luminosity. (But, as we showed above, $D_0$ increases for a compact object.)
However, as CT95 showed, a high accretion rate (${\dot M} \gsim 0.3 {\dot M}_{Edd}$)
will {\it reduce} the temperature of the CENBOL catastrophically, and therefore our assumption of
isothermality of the outflow would breakdown at these high rates. It is to be noted that
in the context of stellar physics, it is shown (Pauldrach et al. 1986) that the temperature
stratification in the outflowing wind has little effect on the mass loss rate. 

Having thus been convinced that isothermality of the outflow, at least up to the
sonic point, is easier to maintain near a black hole, we present in this 
{\it paper} a simple derivation of the ratio of the mass outflow 
rate and mass inflow rate assuming the flow is externally collimated. 
We find that the ratio is a function of the compression ratio of 
the gas at the boundary of the hot, dense, centrifugally supported region. 
We estimate that the outflow rate should generally be less than a few percent if the
outflow is well collimated. In Sect. 3, we find some interesting effects when
the region out to the sonic point of the outflow cools down periodically and causes
periodic change of spectral states. Finally, in Sect. 4, we draw our conclusions.

\section{Derivation of the outflow rate}

Figure 1a shows the schematic nature of the inflow and outflow 
that is understood to be taking place around a black hole.
Rotational motion brakes the flow and forms a dense boundary layer (CENBOL)
around a black hole. Matter compressed and heated (over and
above the heating due to geometric compression) at the CENBOL 
comes out in between the centrifugal barrier and the funnel wall (Molteni et al. 1994, C96).
For present paper, in Fig. 1b, we choose a simplified 
description of this system, where the toroidal  CENBOL is replaced by a 
quasi-spherical one.  Matter is assumed to fall in a conical shape. The 
sub-Keplerian, hot and dense, quasi-spherical region  may also form
due to pair-plasma pressure or pre-heating effects, but the details are
not essential. The outflowing wind is assumed to be 
also conical in shape for simplicity and is flowing out along 
the axis. It is assumed that the wind is collimated by an external pressure.
Both the inflow and the outflow are assumed to be thin 
enough so that the velocity and density variations across the flow could be ignored.

\begin {figure}
\vbox{
\vskip +6.0cm
\hskip 0.0cm
\centerline{
\psfig{figure=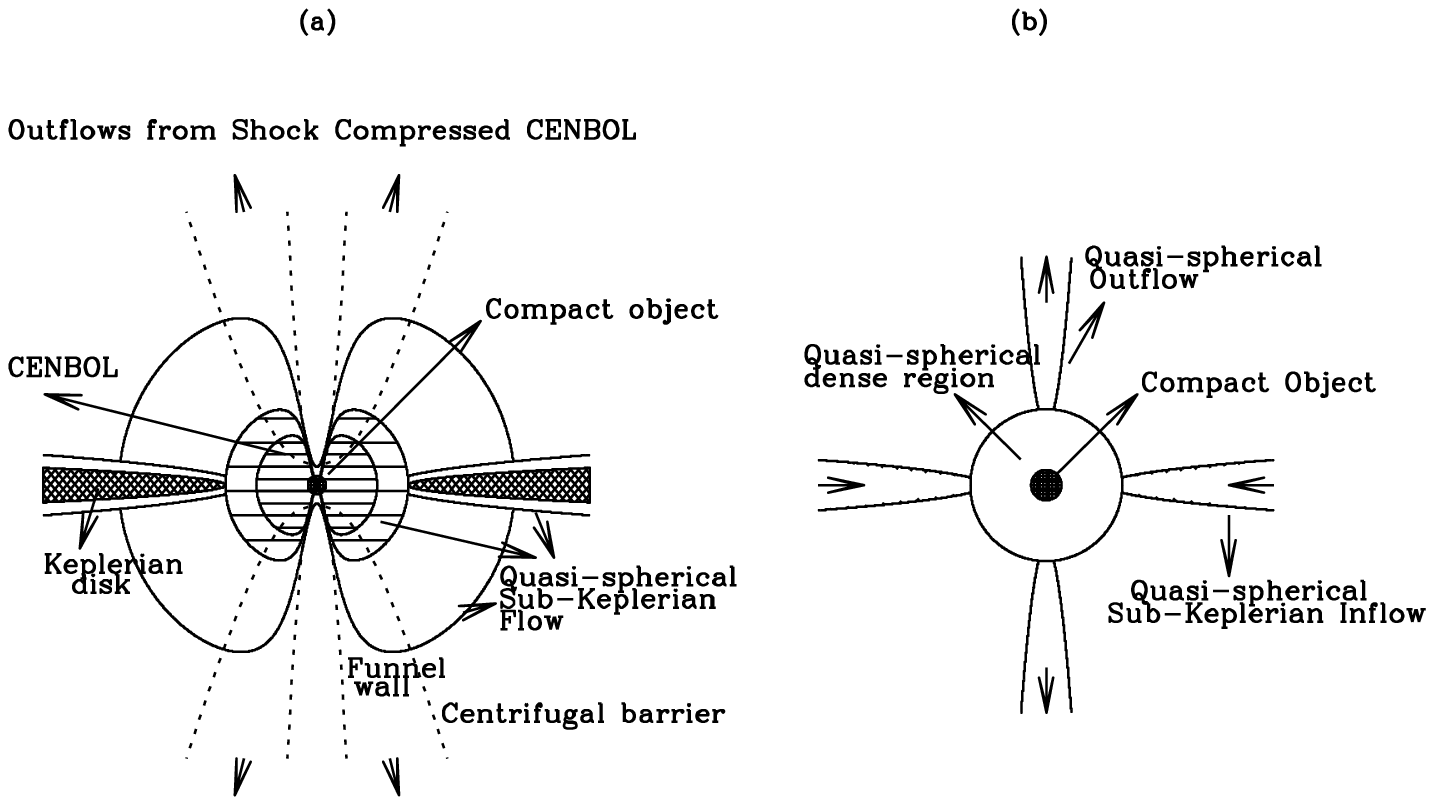,height=10truecm,width=10truecm,angle=0}}}
\vspace{-0.5cm}
\end{figure}
\begin{figure}
\noindent {\small {\bf Fig. (1a).} : 
Schematic diagram of inflow and outflow around a compact object. The hot, dense region around
the object either due to a centrifugal barrier or due to a plasma pressure effect
or pre-heating, acts like a `stellar surface' from which the outflowing wind is developed.
In ({\bf a}), the actual flow behavior is shown. In ({\bf b}), a simplified model is
depicted where the toroidal CENBOL is replaced by a quasi-spherical CENBOL. }
\end{figure}

The accretion rate of the incoming accretion flow is given by,
$$
{\dot M}_{in} = \Theta_{in} \rho \vartheta r^2 .
\eqno{(1)}
$$
Here, $\Theta_{in}$ is the solid angle subtended by the inflow, $\rho$ and
$\vartheta$ are the density and velocity respectively, and $r$ is the
radial distance. For simplicity, we assume geometric units ($G=1=M_{BH}=c$;
$G$ is the gravitational constant, $M_{BH}$ is the mass of the central black hole,
and $c$ is the velocity of light) to measure all the quantities. 
In these units, for a freely falling gas,
$$
\vartheta (r)= [\frac{1-\Gamma}{r}]^{1/2}
\eqno{(2)}
$$
and
$$
\rho(r) = \frac {{\dot M}_{in}}{\Theta_{in}}(1-\Gamma)^{-1/2} r^{-3/2}
\eqno{(3)}
$$
Here, $\Gamma/r^2$ (with $\Gamma$ assumed to be a 
constant) is the outward force due to radiation. 

We assume that the boundary of the denser cloud (say, the shock) is at $r=r_s$
(typically a few to few tens of Schwarzschild radii, see, C90, C96)
where the inflow gas is compressed. The compression could be
abrupt due to a standing shock or gradual as in a shock-free flow
with angular momentum. These details are irrelevant. At this barrier, then 
$$
\rho_+(r_s) = R \rho_- (r_s) 
\eqno{(4a)}
$$
and 
$$
\vartheta_+(r_s) = R^{-1} \vartheta_- (r_s) 
\eqno{(4b)}
$$
where $R$ is the compression ratio. The exact value of the compression ratio
is a function of the flow parameters, such as the specific energy and the
angular momentum (e.g., Chakrabarti 1990 [hereafter C90]).
Here, the subscripts $-$ and $+$ denote the pre-shock and post-shock 
quantities respectively. At the shock surface, the total pressure 
(thermal pressure plus ram pressure) is balanced.
$$
P_- (r_s) + \rho_- (r_s) \vartheta_-^2 (r_s)
= P_+ (r_s) + \rho_+ (r_s) \vartheta_+^2 (r_s).
\eqno{(5)}
$$
Assuming that the thermal pressure of the pre-shock incoming flow is 
negligible compared to the ram pressure, using Eqs. (4a-b) we find,
$$
P_+(r_s) = \frac{R-1}{R} \rho_-(r_s) \vartheta_-^2 (r_s).
\eqno{(6)}
$$
The isothermal sound speed in the post-shock region is then,
$$
C_s^2= \frac{P_+}{\rho_+}=\frac{(R-1)(1-\Gamma)}{R^2}\frac{1}{r_s}
=\frac{(1-\Gamma)}{f_0 r_s}
\eqno{(7)}
$$
where $f_0=R^2/(R-1)$. 
An outflow which is generated from this dense region with very low flow
velocity along the axis is necessarily subsonic in this region,
however, at a large distance, the outflow velocity is expected to be
much higher compared to the sound speed, and therefore the flow must be
supersonic. In the subsonic region of the outflow, the pressure and density
are expected to be almost constant and thus it is customary to 
assume isothermality conditions up to the sonic point. As argued in the
introduction, in the case of black hole accretion also, such an assumption
may be justified. With the isothermality assumption or a given temperature
distribution ($T \propto r^{-\beta}$ with $\beta$ a constant) the result 
is derivable in analytical form. The sonic point conditions are obtained 
from the radial momentum equation, 
$$
\vartheta \frac{d\vartheta}{dr} + \frac{1}{\rho}\frac{dP}{dr} 
+\frac{1-\Gamma}{r^2} = 0 
\eqno{(8)}
$$
and the continuity equation
$$
\frac{1}{r^2}\frac{d (\rho \vartheta r^2)}{dr} =0
\eqno{(9)}
$$
in the usual way, i.e., by eliminating $d\rho/dr$,
$$
\frac{d\vartheta}{dr}= \frac{N}{D}
\eqno{(10)}
$$
where
$$
N=\frac{2 C_s^2}{r} - \frac{1-\Gamma}{r^2}
$$
and
$$
D=\vartheta - \frac{C_s^2}{\vartheta}
$$
and putting $N=0$ and $D=0$. These conditions yield, 
at the sonic point $r=r_c$, for an isothermal flow,
$$
\vartheta (r_c) = C_s 
\eqno{(11a)}
$$
and
$$
r_c = \frac{1-\Gamma}{2 C_s^2}=\frac {f_0 r_s}{2}
\eqno{(11b)}
$$
where we have utilized Eq. (7) to substitute for $C_s$. 

Since the sonic point of a hotter outflow is located closer 
to the black hole, clearly, the condition of isothermality
is best maintained if the temperature is high enough. However if the temperature
is too high, so that $r_c <r_s$, then the flow has to bore a hole through the
cloud just as in the `cauldron' model, although it is a different 
situation --- here the temperature is high, while in the `cauldron' model
the temperature was low. In reality, a pre-defined funnel caused by a 
centrifugal barrier does not require the flow to bore any nozzle at all,
but our simple quasi-spherical calculation fails to describe this case properly.
This is done in detail in Das \& Chakrabarti (1999).

The constancy of the integral of the radial momentum equation 
(Eq. (8)) in an isothermal flow gives: 
$$
C_s^2 ln \ \rho_+ -\frac{1-\Gamma}{r_s} =
\frac{1}{2}C_s^2 + C_s^2 ln \ \rho_c -\frac{1-\Gamma}{r_c}
\eqno{(12)}
$$
where, we have ignored the initial value of the outflowing 
radial velocity $\vartheta (r_s)$ at the dense region boundary, 
and used Eq. (11a). We have also put $\rho(r_c)=\rho_c$ 
and $\rho(r_s) = \rho_+$. After simplification, we obtain,
$$
\rho_c =\rho_+  exp (-f)
\eqno{(13)}
$$
where,
$$
f= f_0 - \frac{3}{2} .
$$
Thus, the outflow rate is given by,
$$
{\dot M}_{out} = \Theta_{out} \rho_c \vartheta_c r_c^2 
\eqno{(14)}
$$
where $\Theta_{out}$ is the solid angle subtended by the outflowing cone. 
Upon substitution, one obtains,
$$
\frac{{\dot M}_{out}} {{\dot M}_{in}} = R_{\dot m}
=\frac{\Theta_{out}}{\Theta_{in}} \frac{R}{4} f_0^{3/2} exp \ (-f)
\eqno{(15)}
$$
which explicitly depends only on the compression ratio:
$$
\frac{{\dot M}_{out}}{{\dot M}_{in}} =R_{\dot m}=
\frac{\Theta_{out}}{\Theta_{in}}\frac{R}{4} 
[\frac{R^2}{R-1}]^{3/2} exp  (\frac{3}{2} - \frac{R^2}{R-1})
\eqno{(16)}
$$
apart from the geometric factors. Notice that this simple result 
does not depend on the location of the sonic point or the
the size of the shock or the outward radiation 
force constant $\Gamma$. This is because the Newtonian potential 
was used throughout and the radiation force was also assumed 
to be very simple minded ($\Gamma/r^2$). Also, 
effects of centrifugal force were ignored. Similarly, the ratio
is independent of the mass accretion rate which should be valid only for
low luminosity objects. For high luminosity flows, Comptonization would
cool the dense region completely (CT95) and the mass loss will be negligible.
Pair plasma supported quasi-spherical shocks form for low luminosity as well
(Kazanas \& Ellison 1986). In reality there would be a dependence on these 
quantities when full general relativistic considerations of the rotating flows are made.

Figures  2a and b contain the basic results. Figure 2a shows the
ratio $R_{\dot m}$ as a function of the compression ratio $R$ (plotted from $1$ to $7$).
Figure 2b shows the same quantity as a function of the polytropic constant 
$n=(\gamma-1)^{-1}$ (drawn from $n=3/2$ to $3$), $\gamma$ being the adiabatic
index. Figure 2a is drawn for any generic compression ratio and Fig. 2b is 
drawn assuming the strong shock limit only: $R=(\gamma+1)/(\gamma-1)=2n+1$. 
In both the curves, $\Theta_{out} \sim \Theta_{in}$ has been assumed 
for simplicity. Note that if the compression (over and above the 
geometric compression) does not take place (namely, if the denser 
region does not form), then there is no outflow in this model. Indeed for, $R=1$, 
the ratio $R_{\dot m}$ is zero as expected. Thus the driving 
force of the outflow is primarily coming from the hot and compressed region.

\begin {figure}
\vbox{
\vskip 3.0cm
\hskip 0.0cm
\centerline{
\psfig{figure=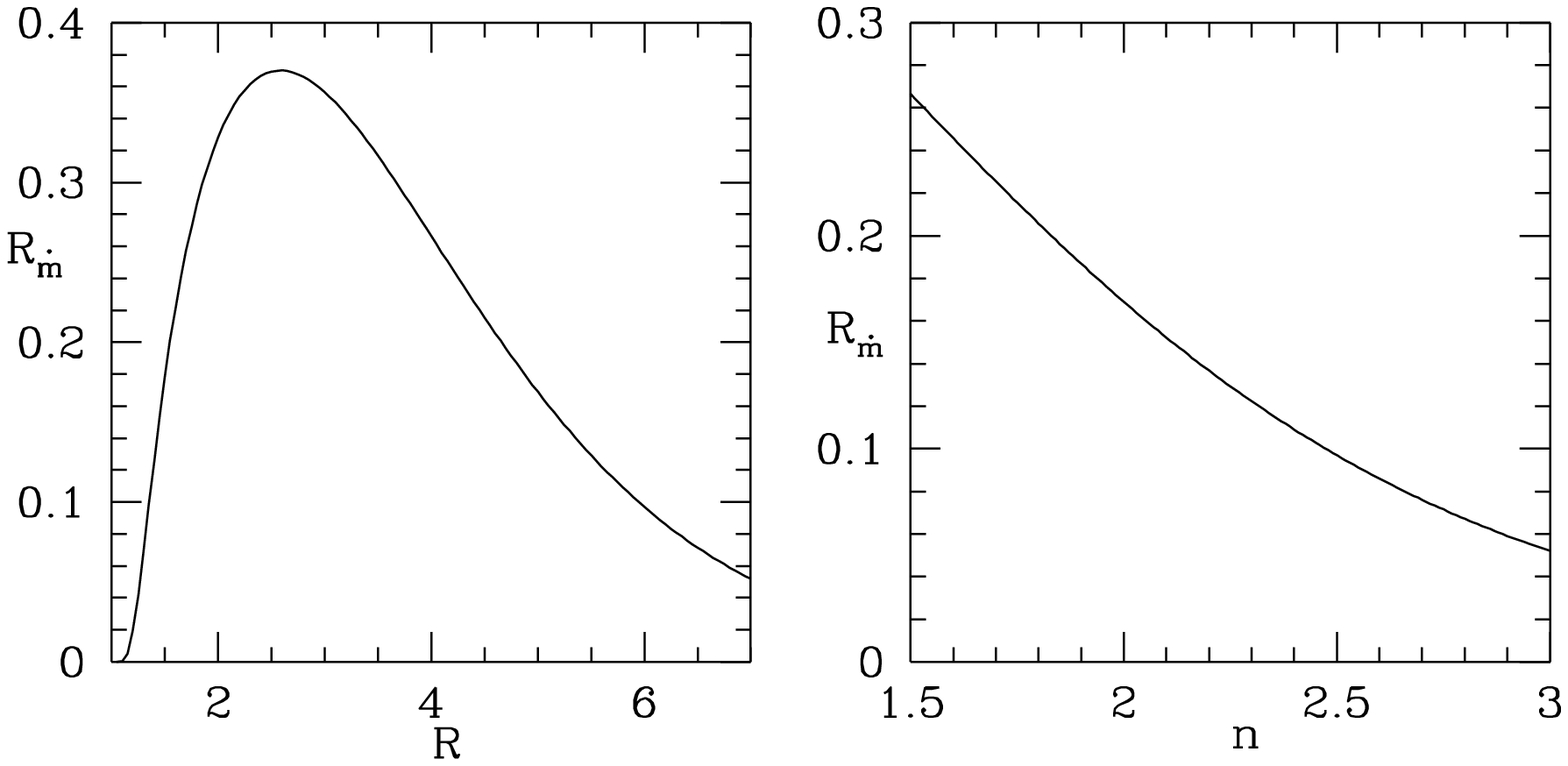,height=10truecm,width=10truecm,angle=0}}}
\vspace{-0.5cm}
\end{figure}
\begin{figure}
\noindent {\small {\bf Fig. 2.} : Ratio ${\dot R}_{\dot m}$ of the outflow rate and the inflow rate
as a function of the compression ratio of the gas at the dense region boundary ({\bf a}). 
In ({\bf b}), variation of the ratio with the polytropic constant $n$ in the strong 
shock limit is shown. Solid angles subtended by the inflow and the outflow are assumed 
to be comparable for simplicity. }
\end{figure}

In a relativistic inflow or for a radiation dominated inflow, $n=3$ and $\gamma=4/3$. 
In the strong shock limit, the compression ratio is $R=7$ and the ratio 
of inflow and outflow rates becomes,
$$
R_{\dot m}=0.052 \ \frac{\Theta_{out}}{\Theta_{in}}.
\eqno{(17a)}
$$
For the inflow of a mono-atomic ionized gas $n=3/2$ and $\gamma=5/3$. 
The compression ratio is $R=4$, and the ratio in this case becomes,
$$
R_{\dot m}=0.266 \ \frac{\Theta_{out}}{\Theta_{in}}.
\eqno{(17b)}
$$
Since $f_0$ is smaller for a $\gamma=5/3$ case, the density at the
sonic point in the outflow is much higher 
(due to the exponential dependence of density on $f_0$, see, eq. 7) 
which causes the higher outflow rate, even when the actual jump in density
in the postshock region, the location of the
sonic point and the velocity of the flow at the sonic point are much lower.
It is to be noted that generally for $\gamma >1.5$ shocks are not
expected (C90), but the centrifugal barrier supported dense region
would still exist. As is clear, the entire behavior of the outflow
depends only on the compression ratio, $R$ and the collimating
property of the outflow $\Theta_{out}/\Theta_{in}$.

Outflows are usually concentrated near the axis, while the inflow is near
the equatorial plane. Assuming a half angle of $10^o$ in each case, 
we obtain,
$$
\Theta_{in}= \frac {2 \pi^2}{9}; \ \ \ \ \ \Theta_{out}= \frac {\pi^3}{162}
$$
and 
$$
\frac{\Theta_{out}}{\Theta_{in}} =\frac{\pi}{36} .
\eqno{(18)}
$$
The ratios of the rates for $\gamma=4/3$ and $\gamma=5/3$ are then
$$
R_{\dot m}=0.0045
\eqno{(19a)}
$$
and 
$$
R_{\dot m}= 0.023
\eqno{(19b)}
$$
respectively. Thus, in quasi-spherical systems, in the case of strong 
shock limit, the outflow rate is at most a few percent of the inflow. 
If this assumption is dropped, then for flow with a weaker shock
the rate could be much higher (see, Fig. 2a).

The angle $\Theta_{out}$ must be related to the collimation property of the
ambient medium as well as the strength of the angular momentum barrier, although
it is doubtful if matter achieves the observed collimation
right close to the black hole. The stronger the
barrier is, the higher is $\Theta_{out}$ (Molteni et al. 1994)
and therefore the higher the loss rate. If $\Theta_{out}$
is sufficiently high and $\Theta_{in}$ is low ($\Theta_{out}+\Theta_{in}=4\pi$)
the outflow may cause a {\it complete} evacuation of the disk
causing a quiescence state of the black hole candidates. The peak area in 
Fig. 2a would have interesting effects on the relation between spectral states and 
quasi-periodic oscillations of spectra of black holes as will be discussed in Sect. 4.

It is to be noted that the above expression for the outflow rate is strictly
valid if the flow could be kept isothermal at least up to the sonic point. If
this assumption is dropped the expression for the outflow rate becomes
dependent on several parameters. As an example, we consider a polytropic
outflow of the same index $\gamma$ but of a different entropy 
function $K$ (we assume the equation of state to be $P=K\rho^\gamma$, with
$\gamma\neq 1$) the expression (11b) would be replaced by
$$
r_c=\frac{f_0r_s}{2\gamma}
\eqno{(20)}
$$
and Eq. (12) would be replaced by
$$
n a_+^2 - \frac{1-\Gamma}{r_s}=(\frac{1}{2} + n) a_s^2 - \frac{1-\Gamma}{r_c}
\eqno{(21)}
$$
where $n=1/(\gamma-1)$ is the polytropic constant of the flow
and $a_+=(\gamma P_+/\rho_+)^{1/2}$ and $a_c=(\gamma P_c/\rho_c)^{1/2}$ 
are the adiabatic sound speeds at the starting point and 
the sonic point of the outflow. It is easily shown that
a power law temperature falloff of the outflow ($T\propto r^{-\beta}$) would yield
$$
R_{\dot m}= \frac{\Theta_{out}}{\Theta_{in}} (\frac{K_i}{K_o})^n 
(\frac{f_0}{2\gamma})^{\frac{3}{2}-\beta},
\eqno{(22)}
$$
where $K_i$ and $K_o$ are the entropy functions of the inflow and the outflow. 
This derivation is strictly valid for a non-isothermal flow. Since $K_i<K_o$, 
$n>3/2$ and $f_0>2\gamma$, for ${\Theta_{out}} \sim {\Theta_{in}}$, $R_{\dot m} <<1$ 
is guaranteed provided $\beta >\frac{3}{2}$, i.e., if the temperature 
falls off sufficiently rapidly. For an isothermal flow $\beta=0$ and the rate 
tends to be higher. Note that since $n \sim \infty$ in this case, any small
jump in entropy due to compression will balance the effect of the $f_0^{-3/2}$ factor.
Thus $R_{\dot m}$ remains smaller than unity. The first factor decreases with the entropy
jump while the second factor will have a minimum at $R=2$ when $\beta<3/2$.
Thus the solution is still expected to be similar to what is shown in Fig. 2a-b.
Numerical results of the transonic flow  using a  non-isothermal equation of state are discussed 
elsewhere (Das \& Chakrabarti, 1999).

\section{Effects of outflows from shocked compressed accretion}

One can discuss the effects of outflows of a stellar black candidate such as GRS 1915+105.
Particularly interesting is that it shows QPO of around 1-15Hz and sometimes flares at around 0.01Hz
(Paul et al. 1998). An interesting property of our solution (Fig. 2a) is that 
the outflow rate is peaked at around $R=2.5$ ($f_0\sim 4$), when the
shock is of `average' strength. On either side, the outflow rate falls off very rapidly
and this may have significant observational effects. As Eq. (16)
is independent of shock location, shocks oscillating with time period
similar to the infall time in $r<r_s$ region (causing a 1-10Hz QPO 
in the hard state (CT95, C96)) will continue to have outflows and gradually fill the 
relatively slowly moving (sonic) sphere of radius $r=r_c=f_0 r_s/2$ till
$<\rho> r_c \sigma_T \sim 1$ when this region would be cooled down
catastrophically by inverse Comptonization. (In general, spectra may soften
in the presence of outflows even in the hard state, see, Chakrabarti, 1998) 
At this stage: (a) $r<r_c$ region 
would be drained to the hole in $t_{fall} \sim r_c^{3/2} 2GM/c^2 = 0.1 (\frac{r_s}{50})^{3/2} 
(\frac{M}{10 M_{\odot}}) (\frac{f_0}{4})$ seconds. This is typically what is observed 
for GRS1915+105. (Yadav et al., 1999). (b) The shock will disappear, 
and a smaller compression ratio ($R\rightarrow 1$) would stop the outflow 
(Fig. 2a). In other words, during burst/quiescence QPO phase, 
the outflow would be blobby. This is also reported (Mirabel \& Rodriguez, 1998).
(c) The black hole will go to a soft state during a short period. 
If the angular momentum is high enough, so that the outflow rate is really high, this brief period of 
soft state may be prolonged to a longer period of tens of seconds depending on how
the centrifugal barrier is removed by viscosity  generated during shock oscillations.
The fact that shock oscillation causes the 1-15Hz QPO is clearly demonstrated by the
fact that there is more power at high energy 
(Fig. 3 of Paul et al. 1998) and most of the high energy
X-ray radiation is emitted in the post-shock region.

Using our solution, it is easy to compute the interval between 
two bursts during which the object is in QPO phase with 
$\nu_{QPO} \sim 1-10$ Hz. The sonic sphere becomes ready for catastrophic cooling in,
$$
t_{burst} = \frac{4/3 \pi r_c^3 <\rho>}{{\dot M}_{out}}\ \  s 
\sim \frac{40 r_s^2 }{R_{\dot m} {\dot M}_{in}}\ \  s.
$$
Here, $<\rho>$ is the average density of the sonic sphere and $<\rho>r_c \sigma_T\sim 1$,
$\sigma_T=0.4$ is the Thomson scattering cross-section. 
For an average shock $R$ around $2.5$, $f_o$ stays close to $4$, and 
with outflow and inflow of roughly equal angular dimension ($\theta \sim 45^o$), 
$\Theta_{out}/\Theta_{in} \sim 0.17$ and $R_{\dot m} \sim  0.06$ (using peak value in Fig. 3b).
Putting typical values of a hard state ${\dot M}=0.1 {\dot M}_{Edd}$, and $r_s=50$, $M=10M_\odot$ 
in the above equation, we obtain,
$$
t_{burst}= 107 (\frac{r_s}{50})^2 (\frac{M}{10M_\odot}) (\frac{{\dot M}}{0.1 {\dot M}_\odot}) \  \  s.
$$
Our choice of $r_s=50$ is not arbitrary. If the viscosity is low, the angular momentum
could be high enough to have a shock at a larger distance (C90) and the oscillation
frequency of QPO for $r_s=50$ is around $6$Hz as in seen in GRS1915+105 (public RXTE archival data
of May 26th, 1997). Our $t_{burst}=107$s is encouraging, since on that day, bursts did repeat 
with $\nu_{burst} \sim 0.01$Hz. However, the system need not remain steady at these 
frequencies due to non-linear processes such as recycling a part of the wind 
back to the accretion disk. Systematic feeding of matter would raise the accretion rate
decreasing the cooling time and increasing the QPO frequency. Similarly, if the specific angular momentum 
remains higher, cool flow may take longer time to be drained,  to the extent 
that the flow may like to stay for a long time in the burst phase as a soft state. 
When the Keplerian rate is actually increased in the inflow (due to a rise in viscosity
at the outer edge of the disk, say), shocks would permanently cool down 
and outflows would be gradually turned off. 

If the shock strength is on the higher side ($R>2.5$), $f_0>>1$, 
the location of the sonic point $r_c$ increases linearly with $f_0$, 
but the average density $<\rho>$  decreases exponentially (Eq. (13)). 
So, $t_{burst} \sim \infty$ and the object will remain in the hard state.
In this phase, hydrodynamic outflow may form continuously as the rate $R_{\dot m}$ does not
go to zero. The QPO frequency $\nu_{QPO}$ is still determined by the infall 
time scale from the shock location. (This is true if the cooling time and the infall time become 
comparable so that QPO forms in the first place.) Only when the 
Keplerian rate is intrinsically increased due to, say, a rise of viscosity at the
outer boundary of the disk, does the shock gets softened and the $t_{burst}$ starts
getting smaller within observable resolution. Thus, 
in this scenario, in the pure soft state, $R\rightarrow 1$
and in pure hard state (with possible QPO) $R \rightarrow 4-7$. In between there is a possibility of 
having both soft (flaring) and hard states (including QPO) switching in tens to hundreds of seconds with
periodicity of $t_{burst}$.

\section{Concluding remarks}

Although the outflows are common in many astrophysical systems 
which include compact objects such as black holes and neutron
stars, it  was difficult to compute the outflow rates
since these objects do not have any intrinsic atmospheres
and outflowing matter has to come out from the 
inflow only. We showed in the present paper, that assuming
the formation of a dense region around these objects (as 
provided by a centrifugal barrier, for instance), it is possible 
to obtain the outflow rate in a compact form with an assumption 
of isothermality of the outflow at least up to the sonic point 
and the ratio thus obtained seems to be quite reasonable. 
Computation of the outflow rate with a non-isothermal outflow 
explicitly depends on several flow parameters. Our primary 
goal in this paper was to obtain the rate as a function of 
the compression ratio of the gas and the geometric quantities. We
show that for a given inflow/outflow configuration, the outflow rate
shows a peak as the shock-compression is increased. We do not concern ourselves
with the collimation mechanism. Since observed jets are generally 
hollow, they must be externally supported (either by ambient medium pressure or 
by magnetic hoop stress). This is assumed here for simplicity.
Our assumption of isothermality of the wind till the sonic point is
based on `experience' borrowed from stellar physics. Momentum deposition from the 
hot photons from the dense cloud, or magnetic heating may or may not isothermalize the 
expanding outflow, depending on accretion rates and covering factors. 
However, it is clear that since the 
solid angle at which photons shine on electrons is close to $4\pi$ (as in a 
narrow funnel wall), and since the number of electrons per photon is much
smaller in a compact region, it may be easier to maintain the isothermality
close to a black hole than near a stellar surface.  

The centrifugal pressure supported region that may be present was found to 
be very useful in  explaining the soft and the hard states 
(CT95), rough agreement with power-law slopes in 
soft states (CT95) as well as the amplitude and frequency of QPO (C96) 
in black hole candidates. Therefore, our reasonable estimate 
of the outflow rate from these considerations further supports the view that
such regions may be common around compact objects. Particularly interesting
is the fact that since the wind here is thermally driven, the outflow
ratio is higher for hotter gas, that is, for a low accretion rate. 
It is obvious that the non-magnetized neutron stars should also 
have the same dense region we discussed here and all the considerations
mentioned here would be equally applicable. 

It is to be noted that although the existence of outflows is well known, 
their rates are not. The only definite candidate whose outflow rate is known with 
any certainty is probably SS433 whose mass outflow rate was estimated to be
${\dot M}_{out} \gsim 1.6 \times 10^{-6}  f^{-1} n_{13}^{-1} D_5^2 M_{\sun} $ yr$^{-1}$
(Watson et al. 1986), where $f$ is the volume filling factor, $n_{13}$ 
is the electron density $n_e$ in units of $10^{13}$ cm$^{-3}$, $D_5$ 
is the distance of SS433 in units of $5$kpc. Considering a central 
black hole of mass $10M_{\sun}$, the Eddington rate is ${\dot M}_{Ed} \sim 
0.2 \times 10^{-7} M_{\sun} $ yr$^{-1}$ and assuming an efficiency 
of conversion of rest mass into gravitational energy $\eta \sim 0.06$, the 
critical rate would be roughly ${\dot M}_{crit} = {\dot M}_{Ed} / \eta \sim
3.2 \times 10^{-7} M_{\sun} $ yr$^{-1}$. Thus, in order to produce the outflow rate
mentioned above even with our highest possible estimated $R_{\dot m}\sim 0.4$ (see
Fig. 2a), one must have ${\dot M}_{in} \sim 12.5 {\dot M}_{crit}$ which is very high
indeed. One possible reason why the above rate might have been over-estimated
would be that below $10^{12}$cm from the central mass (Watson et al. 1986), $n_{13} >>1 $ 
because of the existence of the dense region at the base of the outflow. 

We also discussed the effect of outflows on the spectral states and QPO properties
of black hole candidates, such as GRS1915+105.
The presence of QPO in the hard state, and periodic outbursts in intermediate state,
generally absence of QPO in soft states, switching of states from hard to soft 
states in few seconds (free fall time rather than infall time)
due to the presence of sub-Keplerian matter are generally understood
in this scenario. Since the processes are highly non-linear, more detailed
studies are necessary, but we believe that outflows play a major role
in explaining these properties. In any case, qualitative agreement of the
time scales decidedly prove that sub-Keplerian flows exist in a black hole accretion.
Similar formation of outflows in neutron stars also should explain QPOs, especially
that in neutron stars there could be {\it two}  shocks (one near the hard surface
and the other at a similar distance as the shock around a  black hole; C90, C96).
The only difference is that in neutron stars, the magnetic axis 
could be non-aligned with respect to the spin axis, and 
the outflows on either side of the disk would have an opposite
effect in splitting the QPO frequency 
due to Coriolis force as suggested in other contexts
(Titarchuk et al. 1998).

In numerical simulations the ratio of the outflow and inflow has been computed
in several occasions (Eggum et al. 1985; Molteni et al. 1994). 
Eggum et al. (1985)  found the ratio to be $R_{\dot m} \sim 0.004$ for a 
radiation pressure dominated flow. This is generally comparable with what we found
above (Eq. 19a). In Molteni et al. (1994) the centrifugally driven outflowing wind 
generated a ratio of $R_{\dot m}\sim 0.1$. Here, the angular momentum was present 
in both inflow as well as outflow, and the shock was not very strong. Thus,
the result is again comparable with what we find here. On the other hand, when the 
angular momentum is very high, it was seen that the outflow rate becomes comparable
to the inflow rate. In these simulations, it was seen that the
disk mass changes dramatically, and occasionally evacuating the disk also. This is
possible if $\Theta_{out}$ is very large as in our present model.

This work is partially supported by  `Analytical and Numerical studies
of astrophysical flows around black holes and neutron stars' project funded by Department of
Science and Technology and by `Quasi-Periodic Oscillations 
of Black Hole Candidates' project funded by Indian Space Research Organization.

{}

\end{document}